 \documentclass[prl,twocolumn,citeautoscript,amsmath,amssymb]{revtex4}

\usepackage[dvips]{graphicx}% Include figure files
\usepackage{dcolumn}% Align table columns on decimal point
\usepackage{bm}% bold math
\usepackage{times}% Replaces crummy computer modern font by Times New Roman.

\begin{document}

\title{
All-electron GW calculation for molecules: \\ Ionization energy and electron affinity of conjugated molecules
}

\author{San-Huang Ke}

\affiliation{
Department of Physics, Tongji University, 1239 Siping Road, Shanghai 200092, China
}

%\date{May 22, 2007}% It is always \today, today,

\begin{abstract}
An efficient all-electron G$^0$W$^0$ method and a quasiparticle selfconsistent
GW (QSGW) method for molecules are proposed in the molecular orbital space with the full random phase approximation.
The convergence with basis set is examined.
As an application, the ionization energy ($I$) and electron affinity ($A$) of a series of conjugated molecules
(up to 32 atoms) are calculated and compared to experiment. 
The QSGW result improves the G$^0$W$^0$ result and both of them are in significantly better agreement 
with experimental data than those from Hartree-Fock (HF) and hybrid density functional calculations, especially for $A$. 
The nearly correct energy gap and suppressed self-interaction error by the HF exchange make 
our method a good candidate for investigating electronic and transport properties of molecular systems.
\end{abstract}

%\pacs{73.40.Cg, 72.10.-d, 85.65.+h}
\maketitle

%\section{Introduction}
Understanding and manipulating the electron transport through molecules is the basis of molecular electronics
which provides a promising way for achieving the ultimate miniaturization of electronic components \cite{Heath0343}.
To be enable to calculate accurately the transport properties of molecular devices 
is therefore desirable for the interests of physics and applications. 
In this regard, a standard method which has been extensively used is the
density functional theory (DFT) \cite{Parr89} combined with the nonequilibrium Green function (NEGF) method \cite{Datta95}. 
Despite the success of DFT in describing ground-state electronic properties, 
its application to quantum transport faces much more challenges \cite{Toher05146402,Muralidharan06155410,Ke07201102}.
Intrinsic issues with DFT, such as self-interaction error (SIE) and underestimated energy gap, 
will lead to incorrect molecule-lead charge transfer, incorrect position of the chemical potential in the gap between the highest
occupied molecular orbital (HOMO) and the lowest unoccupied molecular orbital (LUMO) as well as the broadening of the HOMO and LUMO, 
especially for weakly coupled systems \cite{Ke07201102}. Consequently, the molecular conductance will be overestimated,
even by orders of magnitude in some cases, as has been shown by some previous work in literature \cite{Toher05146402,Ke07201102}.
 
A good electronic structure method for a NEGF-based transport calculation should satisfy several conditions: 
(i) nearly correct HOMO and LUMO energies and the gap,
(ii) suppressed SIE, (iii) conserving the number of particle (CNP), and (iv) implemented with localized basis functions.
Physically, even an exact DFT calculation with a local exchange-correlation potential ($V_{xc}$)
cannot give a correct HOMO-LUMO gap due to the lack of the discontinuity. Improvement with a nonlocal $V_{xc}$  
is still under development and facing challenges \cite{Ke07201102,Mori-Sanchez08146401,Cohen08792}. 
Different from DFT, Hartree-Fock (HF) method is SIE free and can usually give a reasonable
HOMO energy (Koopermann's theorem) but lacks electron correlation and cannot predict a reasonable LUMO energy.   
Quasi-particle calculation in the GW approximation 
\cite{Aryasetiawan98237,Onida02601,Rostgaard10085103} may provide a solution 
but its non-selfconsistent version (G$^0$W$^0$) has a critical issue with CNP which has been shown to be important for transport calculations
\cite{Thygesen07091101,Thygesen08115333}. 
It also faces challenging computational effort for large molecular systems.
Additionally, in spite of its very successful applications for bulk semiconductors
\cite{Faleev06226402,Zanolli07245121,Kotani07165106}, 
it accuracy for molecules is still an open problem \cite{Tiago08084311,Palummo09084102,Rostgaard10085103,Kaasbjerg10085102} 
considering the stronger electronic relaxation effect \cite{Kaasbjerg10085102}.
Furthermore, as shown in a recent work \cite{Rostgaard10085103}, the core electrons play an important role in determining accurately
the excitation energies of molecules. Therefore, an efficient {\it ab initio} all-electron GW approach being able to deal with large molecules is desirable.

As an effort towards a more accurate quantum transport calculation, 
in this paper, we first propose an efficient all-electron non-selfconsistent G$^0$W$^0$
method based on the full random phase approximation (RPA) using Gaussian basis functions and a HF input.
To achieve a high computational efficiency, it is implemented in the molecular orbital (MO) space 
with techniques for reducing the error coming from the incompleteness of the basis set.
The correlation self-energy is determined first on the imaginary energy axis and then on the real energy axis using the
analytical continuation approach which was proposed originally for a space-time approach \cite{Rojas951827,Rieger99211}.
The convergence with respect to the size of basis set is examined by calculating He and Be atoms using a series of basis sets.  

To address the issues with the G$^0$W$^0$ method, we further implement a quasiparticle selfconsistent GW (QSGW) method 
\cite{Faleev04126406,Faleev06226402,Kotani07165106},
in which a selfconsistency is performed between the G$^0$W$^0$ and a
quasi-DFT calculation with a nonlocal $V_{xc}$ constructed from
the G$^0$W$^0$ selfenergy. In this way, the converged $V_{xc}$ will largely suppress the SIE since the real (nonlocal) exchange 
is used and is also CNP compliant. Furthermore, as is shown in this work, the quasi-DFT calculation with this $V_{xc}$ 
can give a nearly correct HOMO and LUMO energies, even better than the G$^0$W$^0$ result, 
implying that it will satisfy the four conditions for a good NEGF-based quantum transport calculation. 

As an application of our methods, we investigate the ionization energy ($I$) and electron affinity ($A$) of a series of conjugated molecules.
As is well known, conjugated molecules play the most important roles in molecular electronics because of
their small HOMO-LUMO gap, good molecular conductance, and tunable electronic states. 
The systems studied include acetylene (C$_2$H$_2$), ethylene (C$_2$H$_4$), allene (C$_3$H$_4$),
diacetylene (C$_4$H$_2$), benzene (C$_6$H$_6$), phenylacetylene (C$_8$H$_6$), 
naphthalene (C$_{10}$H$_8$), biphenyl (C$_{12}$H$_{10}$), anthracene (C$_{14}$H$_{10}$), and perylene (C$_{20}$H$_{12}$), 
for which reliable experimental data are available \cite{NIST}.
Comparison to the experimental data shows that our G$^0$W$^0$ and QSGW results improve the HF ones significantly, especially for $A$.
The QSGW improves the G$^0$W$^0$ further and shows a very good agreement between theory and experiment.
This indicates that the all-electron GW calculation can describe very well molecular electronic structures.
The computational efficiency of our methods makes it possible to easily deal with
systems consisting of several tens of atoms from the first principles. The satisfaction of the four conditions 
for quantum transport implies the QSGW-based quasi-DFT approach a good candidate for investigating
electronic and transport properties of molecular devices.

%\section{Theory}

The GW selfenergy is divided into a bare-exchange part and a correlation part: $\Sigma = \Sigma^x + \Sigma^c$.
On the imaginary energy axis the correlation part is
\begin{eqnarray}
&\Sigma^c& (\mathbf{r}, \mathbf{r'}; i\omega) \nonumber\\
&=& -\frac{1}{2\pi}
\int_{-\infty}^{\infty} G(\mathbf{r}, \mathbf{r'},
i(\omega+\omega'))W^c (\mathbf{r}, \mathbf{r'}; i\omega')d\omega',
\end{eqnarray}
where $G$ is the Green function and $W^c \equiv W-v$ is the  
screened Coulomb potential minus the bare Coulomb potential. In the MO space
(the MOs are denoted by $m$, $n$, ...) the Green function simply reads,
\begin{equation}
G_{mn} (i\omega)= \frac{\delta_{mn}}{i\omega-\epsilon_n}
\end{equation}
with $\epsilon_n$ denoting the $n$th eigenvalue.
\begin{eqnarray}
W^c(\mathbf{r}, \mathbf{r'}; i\omega) &=& \left[ \delta(\mathbf{r} - \mathbf{r'}) - P(\mathbf{r}, \mathbf{r'}; i\omega)v \right] ^{-1} v - v,  
\end{eqnarray}
and we define
\begin{eqnarray}
\widetilde{W}^c(\mathbf{r}, \mathbf{r'}; i\omega) &\equiv& \left[ \delta(\mathbf{r} - \mathbf{r'}) - P(\mathbf{r}, \mathbf{r'}; i\omega)v \right] ^{-1} - 1,
\end{eqnarray}
where $P$ is the polarization function which can be determined by RPA:
\begin{equation}
P(\mathbf{r}, \mathbf{r'}; i\omega) = \sum_{kl} (f_k - f_l)
\frac{\psi_k(\mathbf{r})\psi_l^{*}(\mathbf{r})\psi_l(\mathbf{r'})\psi_k^{*}(\mathbf{r'})}{i\omega-(\epsilon_l-\epsilon_k)},
\end{equation}
where $f$ is occupation number and $\psi$'s are the MOs.
If the MO space is complete, $P$ can be expressed as
\begin{equation}
P_{mn} (i\omega) = \sum_{kl} \frac{f_k-f_l}{\epsilon_l-\epsilon_k-i\omega}
O_m^{kl} O_n^{kl}
\end{equation}
in terms of the three-center overlap integrals of the MOs,  
\begin{equation}
O_m^{kl} \equiv \int d^3 \mathbf{r}
\psi_m(\mathbf{r})\psi_k(\mathbf{r})\psi_l(\mathbf{r}).
\end{equation}
However, in practice, the incompleteness of the MO space will cause error in the
calculation of $P$, and even worse, causes larger error in the product of
$Pv$. To suppress this error, we avoid the individual calculations of $P$ and $v$ but calculate the product as a
whole in terms of the electron integrals,
\begin{equation}
\left[ P(i\omega)v \right]_{mn} = \sum_{kl} \frac{f_k-f_l}{\epsilon_l-\epsilon_k-i\omega}
O_m^{kl} C_n^{kl},
\end{equation}
where, 
\begin{equation}
C_n^{kl} \equiv \iint d^3 \mathbf{r} \, d^3 \mathbf{r'} \,
\frac{\psi_n(\mathbf{r})\psi_k(\mathbf{r'})\psi_l(\mathbf{r'})}{|\mathbf{r}-\mathbf{r'}|}
\end{equation}
are the three-center Coulomb integrals of the MOs.
The two kinds of three-center MO integrals are determined in terms of the
corresponding three-center integrals of the atomic orbitals (AO) presented by the
standard contracted Gaussian basis functions, which can then be calculated analytically.
After combining Eqs. (1), (2), (4), and (8), we get 
\begin{eqnarray}
\Sigma^c_{mn} (i\omega) &=& -\frac{1}{2\pi} \sum_{jkl} C^{kj}_m O^{jl}_n
\int_{-\infty}^{\infty} \nonumber
\frac{\widetilde{W}^c_{kl}(\omega')}{i(\omega+\omega')-\epsilon_j} d\omega' \\
&\equiv& -\frac{1}{2\pi} \sum_{jkl} C^{jk}_m \mathcal{W}_{jkl} O^{jl}_n
\end{eqnarray}
which is simply a product of the three three-dimensional arrays. The integral $\mathcal{W}_{jkl}$ 
can be calculated by Gaussian quadrature. Because $\widetilde{W}^c$ varies extreme smoothly 
on the imaginary energy axis, only very few energy
grids (about 100 points from 0 to $\infty$) are needed to well converge the elements of $\mathcal{W}$.

After having obtained $\Sigma^c$ on the imaginary energy axis, we determine its values on the real energy axis 
using the analytical continuation approach \cite{Rojas951827,Rieger99211}, and least-square fit each element to the multipole form  
using a BFGS optimizaton technique, 
$a_0 + \sum_{i=1}^n \frac{a_i}{\omega - b_i}$
with complex parameters $a_i$ and $b_i$. For bulk semiconductors it was found that  
$n=2$ is usually good enough \cite{Rojas951827,Rieger99211}. For molecules, however,  
we find that $n=6$ is generally required to give an extremely stable and accurate fitting. 

To address the issues with the G$^0$W$^0$ method, we implement a QSGW approach, in which the G$^0$W$^0$ selfenergy is used 
to construct a nonlocal exchange-correlation potential $V_{xc}$ for a quasi-DFT calculation.
In practice, we follow the scheme proposed in Ref.\cite{Faleev04126406,Faleev06226402,Kotani07165106}, 
which has been shown to be very successful for bulk semiconductors, 
but now implemented in the MO space for molecules,
\begin{equation}
[V_{xc}]_{mn} = \mathrm{Re} \left[ \Sigma^c (\epsilon_m)+\Sigma^c (\epsilon_n) \right]_{mn} /2 + \Sigma^x_{mn}
\end{equation}
with $\Sigma^x$ being calculated in terms of the HF exchange operator. 
The quasi-DFT calculation will generate new eigenvalues and MOs which can be used for the next G$^0$W$^0$ calculation.
This procedure is going on until the maximum change in the $V_{xc}$ elements is smaller than the convergence criteria. 
Previously, the QSGW approach implemented in plane-wave pseudopotential formalism was  
applied to bulk semiconductors and improved significantly the G$^0$W$^0$ band gaps \cite{Faleev04126406,Faleev06226402}.
For molecules, its applicability and quality is still unknown. 
%partly due to the demanding computational effort of the G$^0$W$^0$ method itself.

%\section{Numerical results}
%\subsection{Convergency of the method: He and Be atoms}

Like other correlated electronic structure methods (for instance, 
the second-order M{\o}ller-Plesset theory (MP2) and coupled-cluster method) the G$^0$W$^0$ method 
is more sensitive to the size of basis set than DFT and HF method.
So far, careful examinations for its convergence behavior are still lacking in literature \cite{Rostgaard10085103},
probably leading to the scattering results from different calculations and biased conclusions for different systems.
To examine the basis-set convergence of our method, 
we first calculate the -HOMO energy (i.e., $I$) of He and Be atoms,
adopting different basis sets \cite{NWChem}: 6-31G*, 6-311G**, cc-pvDZ, cc-pvTZ, cc-pvQZ, aug-cc-pvQZ, cc-pv5Z, cc-pv6Z, and aug-cc-pv6Z,
and plot the results in Fig. 1.

\begin{figure}[t]
\includegraphics[width=6.0cm,clip]{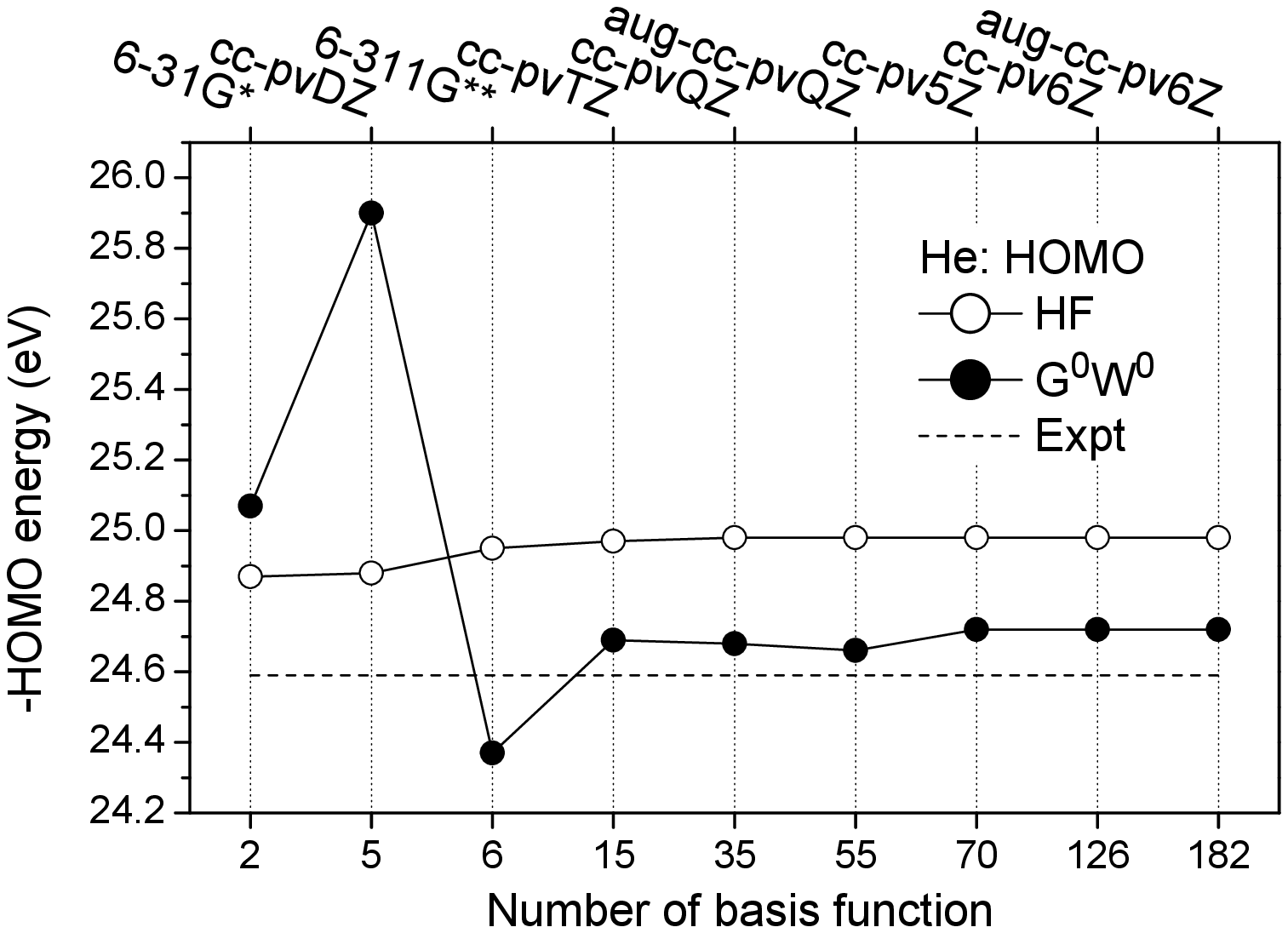}
\includegraphics[width=6.0cm,clip]{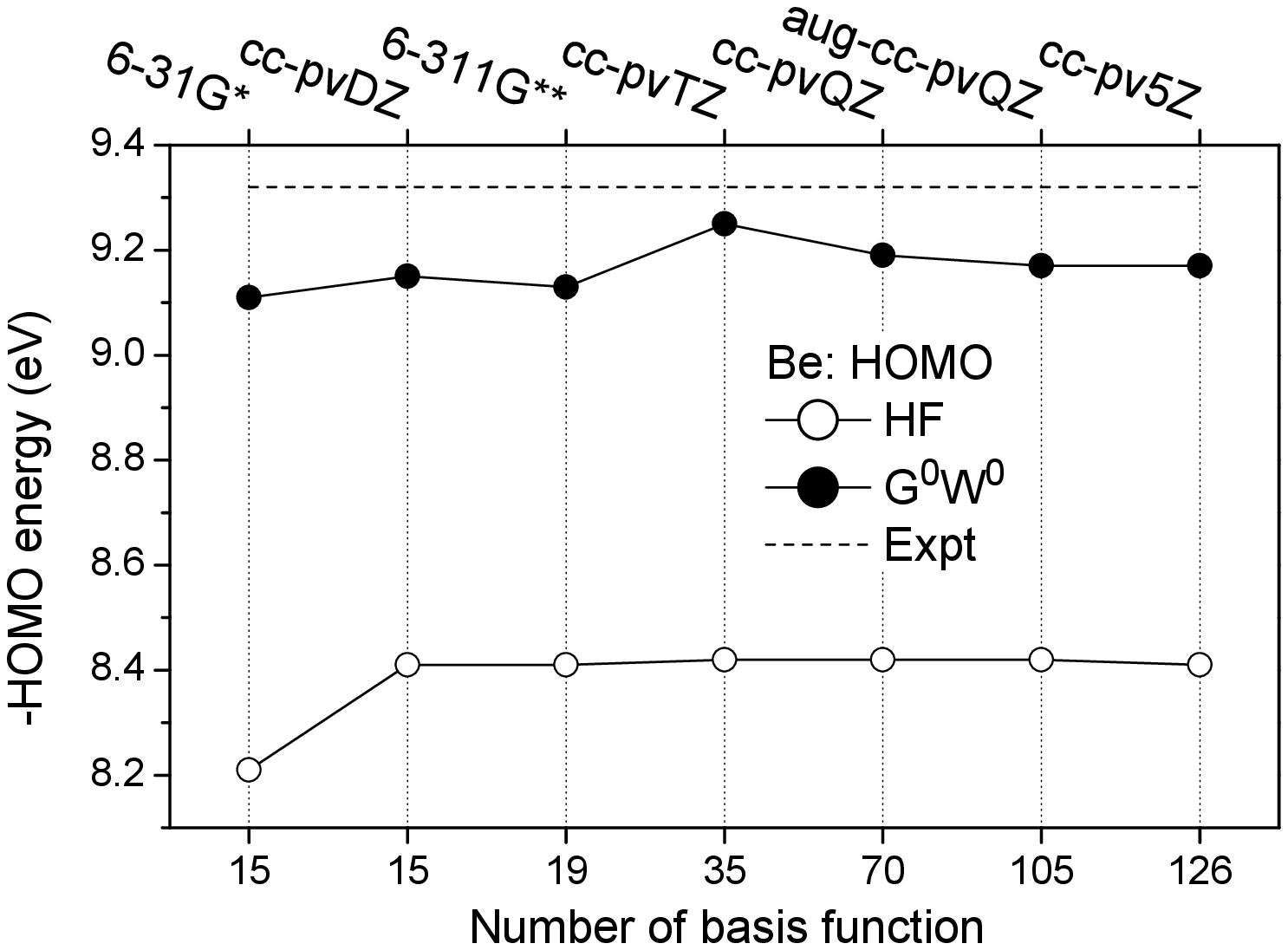}
\caption{\label{fig:atom}
Dependence of the -HOMO energy of He atom (upper panel) and Be atom (lower panel) on the size of basis
set used in the HF and G$^0$W$^0$ calculations.
The experimental result is shown by a horizontal dashed line.
}
\end{figure}

\begin{table*}[bth]
\caption{Calculated (cc-pVTZ level) and experimental results of $I$ and $A$ (in eV) of ten conjugated molecules, as well as their charge-induced
relaxation energies $\Delta E^{+}$ and $\Delta E^{-}$. The experimental data are cited from Ref.\cite{NIST}.}
\label{result}
\begin{center}
\begin{tabular}{ldddddddddddddddd}
\hline
&                & \multicolumn{7}{c}{$I$} & & \multicolumn{7}{c}{$A$} \\
\cline{3-9} \cline{11-17}
& \multicolumn{1}{c}{${\Delta}E^{+}$} & \multicolumn{1}{c}{G$^0$W$^0$} & \multicolumn{1}{c}{QSGW} & \multicolumn{1}{c}{HF} 
& \multicolumn{1}{c}{$\Delta$HF} & \multicolumn{1}{c}{B3LYP} & \multicolumn{1}{c}{$\Delta$B3LYP} & \multicolumn{1}{c}{Expt.}
& \multicolumn{1}{c}{${\Delta}E^{-}$} & \multicolumn{1}{c}{G$^0$W$^0$} & \multicolumn{1}{c}{QSGW} & \multicolumn{1}{c}{HF}
& \multicolumn{1}{c}{$\Delta$HF} & \multicolumn{1}{c}{B3LYP} & \multicolumn{1}{c}{$\Delta$B3LYP} & \multicolumn{1}{c}{Expt.} \\
\hline
%                   dE+    G0W0    QSGW     HF     dHF     B3lpy   dB3lpy  expt    dE-    G0W0    QSGW     HF     dHF      B3lpy   dB3lpy  expt
C$_2$H$_2$        & 0.11 & 11.44 & 11.31 & 11.07 &  9.69 &  8.05 & 11.26 & 11.40 &      &       &       &       &       &       &       &       \\
C$_2$H$_4$        & 0.19 & 10.50 & 10.53 & 10.10 &  8.74 &  6.60 & 10.39 & 10.51 &      &       &       &       &       &       &       &       \\
C$_3$H$_4$        & 0.57 &  9.95 &  9.78 &  9.74 &  8.24 &  6.91 &  9.51 &  9.69 &      &       &       &       &       &       &       &       \\
C$_4$H$_2$        & 0.09 & 10.34 & 10.12 &  9.98 &  8.72 &  7.35 &  9.85 & 10.17 &      &       &       &       &       &       &       &       \\
C$_6$H$_6$        & 0.14 &  9.28 &  9.20 &  9.00 &  7.73 &  6.90 &  9.11 &  9.24 &      &       &       &       &       &       &       &       \\
C$_8$H$_6$        & 0.12 &  8.94 &  8.66 &  8.60 &  8.50 &  6.54 &  8.50 &  8.82 &      &       &       &       &       &       &       &       \\
C$_{10}$H$_8$     & 0.09 &  8.16 &  8.07 &  7.78 &  6.82 &  6.02 &  7.85 &  8.14 & 0.12 & -0.60 & -0.34 & -1.35 & -2.17 &  1.46 & -0.32 & -0.20 \\
C$_{12}$H$_{10}$  & 0.30 &  8.31 &  8.21 &  8.03 &  6.77 &  6.16 &  7.81 &  8.16 & 0.44 & -0.66 & -0.29 & -1.38 & -2.30 &  1.39 & -0.36 &  0.13 \\
C$_{14}$H$_{10}$  & 0.07 &  7.42 &  7.27 &  6.97 &  6.04 &  5.47 &  7.04 &  7.44 & 0.10 &  0.23 &  0.37 & -0.51 & -1.33 &  2.08 &  0.42 &  0.53 \\
C$_{20}$H$_{12}$  & 0.08 &  6.94 &  6.77 &  6.50 &  5.19 &  5.17 &  6.58 &  6.96 & 0.09 &  0.65 &  0.91 &  0.13 & -0.90 &  2.33 &  0.84 &  0.97 \\

\hline
\end{tabular}
\end{center}
\end{table*}

For He atom, the 6-31G*, cc-pvDZ, and 6-311G** basis sets have only 
2, 5, and 6 basis functions, respectively. For such small basis sets its  G$^0$W$^0$ -HOMO energy 
fluctuates remarkably. However, once the basis
set is larger than 15 basis functions the results of both He and Be atoms
tend to be well converged and the fluctuation is around 0.1eV. The important
thing is that for both atoms the G$^0$W$^0$ result improves
significantly the HF result: For He the -HOMO energy is corrected downwards while 
for Be it is corrected upwards, becoming much closer to the experimental data. 

%discussion

%\subsection{Ionization energy and electron affinity of molecules}

For a molecule, its $I$ or $A$ includes two contributions, i.e., the vertical $I$ or $A$ and the corresponding 
charge-induced structural relaxation energy ($\Delta E^{+}$ or $\Delta E^{-}$): 
$I$ = $I_{\rm vert.}$ - $\Delta E^{+}$ and $A$ = $A_{\rm vert.}$ + $\Delta E^{-}$.
$\Delta E^{+,-}$ can be determined by performing a structural relaxation calculation with the charge. 
For the ten conjugated molecules studied in this work we first optimize their 
structures on the DFT/B3LYP/cc-pvTZ level, 
and then calculate their $\Delta E^{+}$ and $\Delta E^{-}$ by allowing further relaxation with the charge.
From the results listed in Table 1 one can see that $\Delta E^{+}$ is comparable with $\Delta E^{-}$
and their values depend strongly on the molecular structure.

%\section{$IE$ and $EA$ of conjugated molecules}

The results of $I$ and $A$ of the ten conjugated molecules given by the different methods are listed in Table 1 together with 
the available experimental data cited from Ref.\cite{NIST}. For $A$ only the four largest molecules are studied, which have a positive or nearly 
zero experimental value. 
%The other molecules are too small to have a positive value of EA because the added electron is unbund, and therefore are not studied. 
The results are also ploted in Fig. 2 for a better vision and comparison.
For molecular systems the hybrid DFT/B3LPY has been proven to be a significant improvement to
DFT/LDA and DFT/GGA because the HF exchange is partly included, suppressing partly the SIE with LDA and GGA \cite{Ke07201102}.
However, as shown in Table 1 and Fig.2, its results of $I$ and $A$ of the conjugated molecules are still far away from being correct.  
In spite of lacking electron correlation, HF method gives much better result of $I$ due to the Koopermann's theorem, but its result of $A$ is 
very bad, being comparable to the B3LYP result but with an opposite trend (see Fig.2).
The total-energy difference calculation using HF ($\Delta$HF) improves HF for $A$ considerably 
but $\Delta$HF result of $I$ becomes much worse because of the inaccurate total energy.
On the other hand, the total-energy difference calculation using B3LYP ($\Delta$B3LPY) gives very good results 
of both $I$ and $A$ because of its accurate total energy.

The G$^0$W$^0$ result improves the HF result considerably for $I$ while significantly for $A$, showing that the correlation 
is more important for the excited states. The QSGW result improves the G$^0$W$^0$ result further, 
being in very good agreement with the available experimental data \cite{NIST} and the $\Delta$B3LPY result. 
This finding is consistent with previous calculations \cite{Faleev04126406,Faleev06226402,Kotani07165106}
which showed a significant improvement from QSGW to G$^0$W$^0$ for band gaps in bulk semiconductors.
We note that the quasi-DFT calculation with the nonlocal $V_{xc}$ from the QSGW method can be a good candidate for NEGF-based quantum transport calculations
considering the facts that (i) it can give nearly correct HOMO and LUMO energies of molecules, 
(ii) it is CNP compliant, (iii) SIE is largely avoided because of the use of the HF exchange, 
and (iv) it is implemented with the local Gaussian basis functions. 

Finally, our work shows that the all-electron {\it ab initio} G$^0$W$^0$ and QSGW methods can describe 
very well the electronic structures of conjugated molecules. This finding is consistent with a previous {\it ab initio} G$^0$W$^0$
calculation for small molecules \cite{Rostgaard10085103}, which also found that the inclusion of 
core-valence exchange is very important in obtaining accurate excitation energies,  but is somewhat different from a recent $\pi$-only model 
GW calculation \cite{Kaasbjerg10085102} which shows that the accuracy of GW method is largely affected 
by the electron-removal and electron-addition induced relaxation.
This difference may be due to the role played by the $\sigma$ states which have much broader energy range, and the resulting $\pi$-$\sigma$ relaxation.

\begin{figure*}[t]
\includegraphics[width=6.1cm,clip]{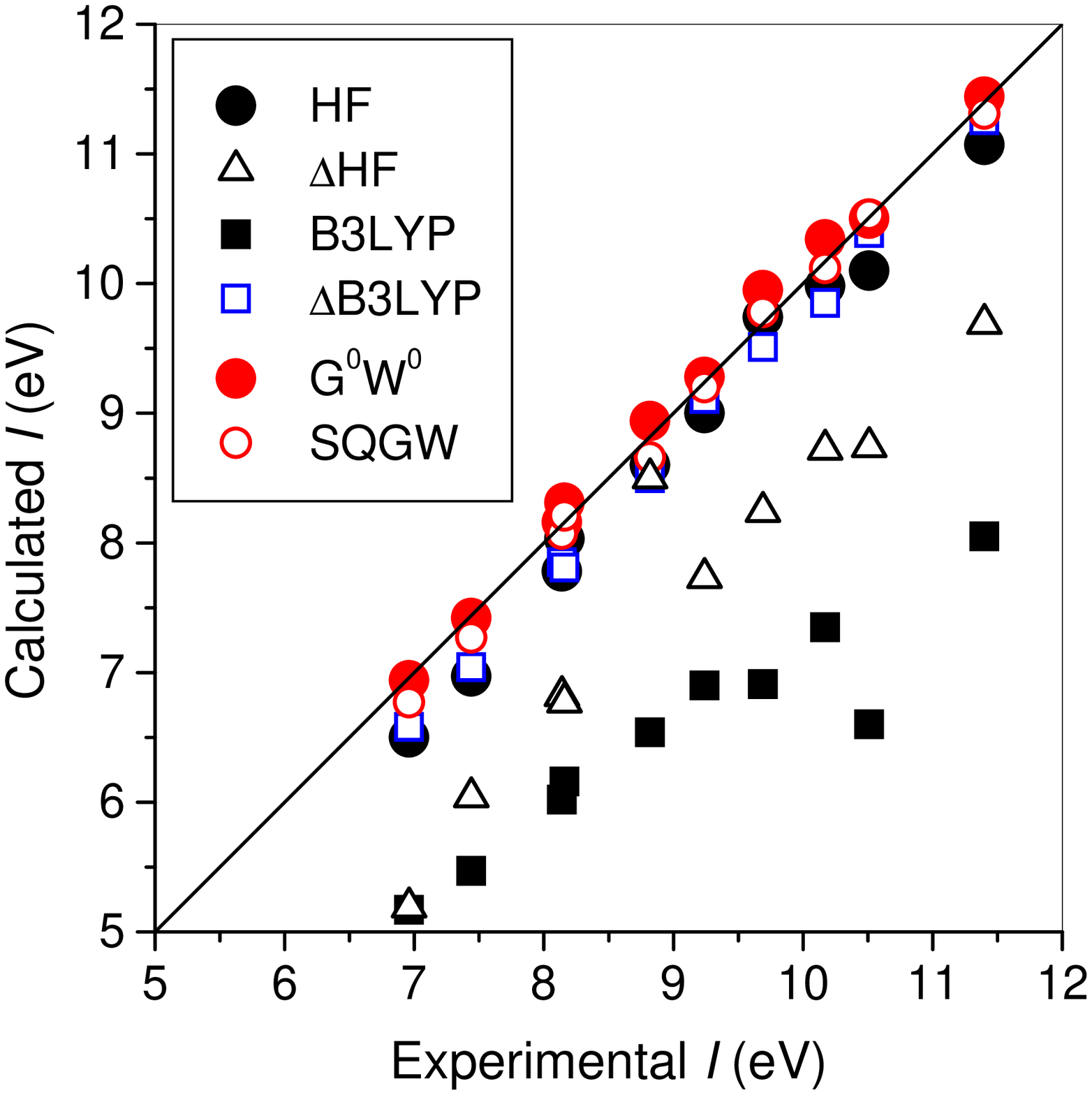}
\includegraphics[width=6.0cm,clip]{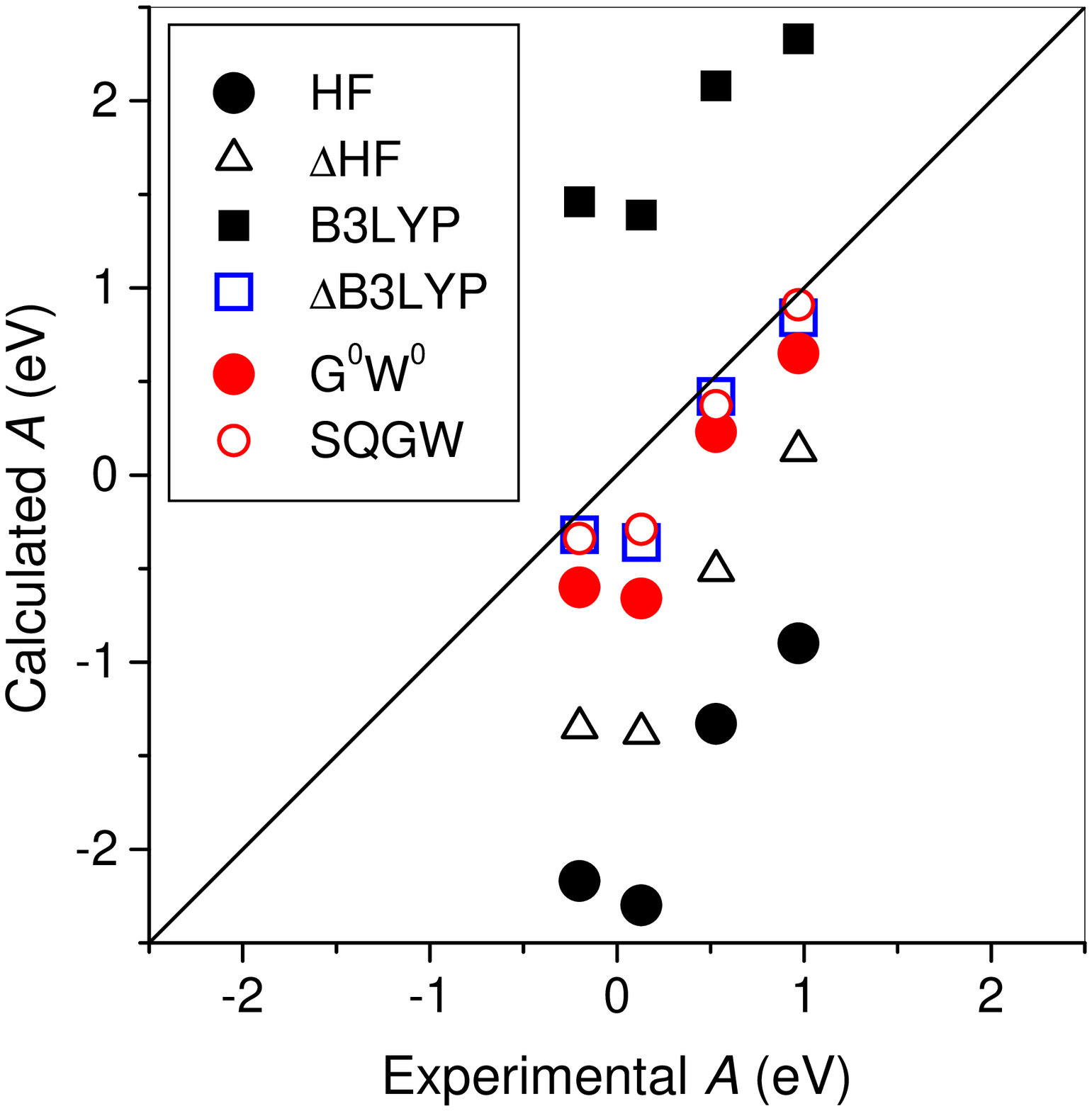}
\caption{\label{fig:m10}
Comparison between the calculated results and experimental data of the $I$ and $A$ of the ten conjugated molecules in Table 1.
}
\end{figure*}

%Summary
%
%In summary, we have proposed a fast nonselfconsistent G$^0$W$^0$ method which is based on the full RPA and 
%implemented in molecular orbital space with techniques for reducing the error coming from the incompleteness of the basis set.
%Based on this, we further implemented a selfconsistent QSGW approach with localized Gaussian basis functions. 
%%
%The high computational efficiency allow us to deal with large molecules and we apply our method to calculate the $I$ and $A$ of ten conjugated 
%molecules with the largest one containing 32 atoms. Our G$^0$W$^0$ method improve HF method significantly, especially for $A$,
%and furthermore, the QSGW improves the G$^0$W$^0$ and gives results of both $I$ and $A$ in very good agreement
%with the available experimental data. 
%%
%The nonlocal exchange-correlation potential created by the QSGW method satisfy the required conditions for NEGF-based
%quantum transport calculations, and therefore can be a good candidate for investigating the transport properties of molecular devices.   

This work was supported by Shanghai Pujiang Program under Grant 10PJ1410000 and
the MOST 973 Project 2011CB922200. 

%\bibliography{me}

\begin{thebibliography}{24}
\expandafter\ifx\csname natexlab\endcsname\relax\def\natexlab#1{#1}\fi
\expandafter\ifx\csname bibnamefont\endcsname\relax
  \def\bibnamefont#1{#1}\fi
\expandafter\ifx\csname bibfnamefont\endcsname\relax
  \def\bibfnamefont#1{#1}\fi
\expandafter\ifx\csname citenamefont\endcsname\relax
  \def\citenamefont#1{#1}\fi
\expandafter\ifx\csname url\endcsname\relax
  \def\url#1{\texttt{#1}}\fi
\expandafter\ifx\csname urlprefix\endcsname\relax\def\urlprefix{URL }\fi
\providecommand{\bibinfo}[2]{#2}
\providecommand{\eprint}[2][]{\url{#2}}

\bibitem[{\citenamefont{Heath and Ratner}(May 2003)}]{Heath0343}
\bibinfo{author}{\bibfnamefont{J.~R.} \bibnamefont{Heath}} \bibnamefont{and}
  \bibinfo{author}{\bibfnamefont{M.~A.} \bibnamefont{Ratner}},
  \bibinfo{journal}{Physics Today} \textbf{\bibinfo{volume}{56(5)}},
  \bibinfo{pages}{43} (\bibinfo{year}{May 2003}).

\bibitem[{\citenamefont{Parr and Yang}(1989)}]{Parr89}
\bibinfo{author}{\bibfnamefont{R.~G.} \bibnamefont{Parr}} \bibnamefont{and}
  \bibinfo{author}{\bibfnamefont{W.}~\bibnamefont{Yang}},
  \emph{\bibinfo{title}{Density-Functional Theory of Atoms and Molecules}}
  (\bibinfo{publisher}{Oxford University Press}, \bibinfo{address}{New York},
  \bibinfo{year}{1989}).

\bibitem[{\citenamefont{Datta}(1995)}]{Datta95}
\bibinfo{author}{\bibfnamefont{S.}~\bibnamefont{Datta}},
  \emph{\bibinfo{title}{Electronic Transport in Mesoscopic Systems}}
  (\bibinfo{publisher}{Cambridge University Press},
  \bibinfo{address}{Cambridge, England}, \bibinfo{year}{1995}).

\bibitem[{\citenamefont{Toher et~al.}(2005)\citenamefont{Toher, Filippetti,
  Sanvito, and Burke}}]{Toher05146402}
\bibinfo{author}{\bibfnamefont{C.}~\bibnamefont{Toher}},
  \bibinfo{author}{\bibfnamefont{A.}~\bibnamefont{Filippetti}},
  \bibinfo{author}{\bibfnamefont{S.}~\bibnamefont{Sanvito}}, \bibnamefont{and}
  \bibinfo{author}{\bibfnamefont{K.}~\bibnamefont{Burke}},
  \bibinfo{journal}{Phys. Rev. Lett.} \textbf{\bibinfo{volume}{95}},
  \bibinfo{pages}{146402} (\bibinfo{year}{2005}).

\bibitem[{\citenamefont{Muralidharan et~al.}(2006)\citenamefont{Muralidharan,
  Ghosh, and Datta}}]{Muralidharan06155410}
\bibinfo{author}{\bibfnamefont{B.}~\bibnamefont{Muralidharan}},
  \bibinfo{author}{\bibfnamefont{A.~W.} \bibnamefont{Ghosh}}, \bibnamefont{and}
  \bibinfo{author}{\bibfnamefont{S.}~\bibnamefont{Datta}},
  \bibinfo{journal}{Phys. Rev. B} \textbf{\bibinfo{volume}{73}},
  \bibinfo{pages}{155410} (\bibinfo{year}{2006}).

\bibitem[{\citenamefont{Ke et~al.}(2007)\citenamefont{Ke, Baranger, and
  Yang}}]{Ke07201102}
\bibinfo{author}{\bibfnamefont{S.-H.} \bibnamefont{Ke}},
  \bibinfo{author}{\bibfnamefont{H.~U.} \bibnamefont{Baranger}},
  \bibnamefont{and} \bibinfo{author}{\bibfnamefont{W.}~\bibnamefont{Yang}},
  \bibinfo{journal}{J. Chem. Phys.} \textbf{\bibinfo{volume}{126}},
  \bibinfo{pages}{201102} (\bibinfo{year}{2007}).

\bibitem[{\citenamefont{Mori-S\'{a}nchez
  et~al.}(2008)\citenamefont{Mori-S\'{a}nchez, Cohen, and
  Yang}}]{Mori-Sanchez08146401}
\bibinfo{author}{\bibfnamefont{P.}~\bibnamefont{Mori-S\'{a}nchez}},
  \bibinfo{author}{\bibfnamefont{A.~J.} \bibnamefont{Cohen}}, \bibnamefont{and}
  \bibinfo{author}{\bibfnamefont{W.}~\bibnamefont{Yang}},
  \bibinfo{journal}{Phys. Rev. Lett.} \textbf{\bibinfo{volume}{100}},
  \bibinfo{pages}{146401} (\bibinfo{year}{2008}).

\bibitem[{\citenamefont{Cohen et~al.}(2008)\citenamefont{Cohen,
  Mori-S\'{a}nchez, and Yang}}]{Cohen08792}
\bibinfo{author}{\bibfnamefont{A.~J.} \bibnamefont{Cohen}},
  \bibinfo{author}{\bibfnamefont{P.}~\bibnamefont{Mori-S\'{a}nchez}},
  \bibnamefont{and} \bibinfo{author}{\bibfnamefont{W.}~\bibnamefont{Yang}},
  \bibinfo{journal}{Science} \textbf{\bibinfo{volume}{321}},
  \bibinfo{pages}{792} (\bibinfo{year}{2008}).

\bibitem[{\citenamefont{Aryasetiawan and Gunnarsson}(1998)}]{Aryasetiawan98237}
\bibinfo{author}{\bibfnamefont{F.}~\bibnamefont{Aryasetiawan}}
  \bibnamefont{and}
  \bibinfo{author}{\bibfnamefont{O.}~\bibnamefont{Gunnarsson}},
  \bibinfo{journal}{Rep. Prog. Phys.} \textbf{\bibinfo{volume}{61}},
  \bibinfo{pages}{237 } (\bibinfo{year}{1998}).

\bibitem[{\citenamefont{Onida et~al.}(2002)\citenamefont{Onida, Reining, and
  Rubio}}]{Onida02601}
\bibinfo{author}{\bibfnamefont{G.}~\bibnamefont{Onida}},
  \bibinfo{author}{\bibfnamefont{L.}~\bibnamefont{Reining}}, \bibnamefont{and}
  \bibinfo{author}{\bibfnamefont{A.}~\bibnamefont{Rubio}},
  \bibinfo{journal}{Rev. Mod. Phys.} \textbf{\bibinfo{volume}{74}},
  \bibinfo{pages}{601} (\bibinfo{year}{2002}).

\bibitem[{\citenamefont{Rostgaard et~al.}(2010)\citenamefont{Rostgaard,
  Jacobsen, and Thygesen}}]{Rostgaard10085103}
\bibinfo{author}{\bibfnamefont{C.}~\bibnamefont{Rostgaard}},
  \bibinfo{author}{\bibfnamefont{K.~W.} \bibnamefont{Jacobsen}},
  \bibnamefont{and} \bibinfo{author}{\bibfnamefont{K.~S.}
  \bibnamefont{Thygesen}}, \bibinfo{journal}{Phys. Rev. B}
  \textbf{\bibinfo{volume}{81}}, \bibinfo{pages}{085103}
  (\bibinfo{year}{2010}).

\bibitem[{\citenamefont{Thygesen and Rubio}(2007)}]{Thygesen07091101}
\bibinfo{author}{\bibfnamefont{K.~S.} \bibnamefont{Thygesen}} \bibnamefont{and}
  \bibinfo{author}{\bibfnamefont{A.}~\bibnamefont{Rubio}}, \bibinfo{journal}{J.
  Chem. Phys.} \textbf{\bibinfo{volume}{126}}, \bibinfo{pages}{091101}
  (\bibinfo{year}{2007}).

\bibitem[{\citenamefont{Thygesen and Rubio}(2008)}]{Thygesen08115333}
\bibinfo{author}{\bibfnamefont{K.~S.} \bibnamefont{Thygesen}} \bibnamefont{and}
  \bibinfo{author}{\bibfnamefont{A.}~\bibnamefont{Rubio}},
  \bibinfo{journal}{Phys. Rev. B} \textbf{\bibinfo{volume}{77}},
  \bibinfo{pages}{115333} (\bibinfo{year}{2008}).

\bibitem[{\citenamefont{v.~Schilfgaarde
  et~al.}(2006)\citenamefont{v.~Schilfgaarde, Kotani, and
  Faleev}}]{Faleev06226402}
\bibinfo{author}{\bibfnamefont{M.}~\bibnamefont{v.~Schilfgaarde}},
  \bibinfo{author}{\bibfnamefont{T.}~\bibnamefont{Kotani}}, \bibnamefont{and}
  \bibinfo{author}{\bibfnamefont{S.}~\bibnamefont{Faleev}},
  \bibinfo{journal}{Phys. Rev. Lett.} \textbf{\bibinfo{volume}{96}},
  \bibinfo{pages}{226402} (\bibinfo{year}{2006}).

\bibitem[{\citenamefont{Zanolli et~al.}(2007)\citenamefont{Zanolli, Fuchs,
  Furthmuller, v.~Barth, and Bechstedt}}]{Zanolli07245121}
\bibinfo{author}{\bibfnamefont{Z.}~\bibnamefont{Zanolli}},
  \bibinfo{author}{\bibfnamefont{F.}~\bibnamefont{Fuchs}},
  \bibinfo{author}{\bibfnamefont{J.}~\bibnamefont{Furthmuller}},
  \bibinfo{author}{\bibfnamefont{U.}~\bibnamefont{v.~Barth}}, \bibnamefont{and}
  \bibinfo{author}{\bibfnamefont{F.}~\bibnamefont{Bechstedt}},
  \bibinfo{journal}{Phys. Rev. B} \textbf{\bibinfo{volume}{75}},
  \bibinfo{pages}{245121} (\bibinfo{year}{2007}).

\bibitem[{\citenamefont{Kotani and v.~Schilfgaarde}(2007)}]{Kotani07165106}
\bibinfo{author}{\bibfnamefont{T.}~\bibnamefont{Kotani}} \bibnamefont{and}
  \bibinfo{author}{\bibfnamefont{M.}~\bibnamefont{v.~Schilfgaarde}},
  \bibinfo{journal}{Phys. Rev. B} \textbf{\bibinfo{volume}{76}},
  \bibinfo{pages}{165106} (\bibinfo{year}{2007}).

\bibitem[{\citenamefont{Kaasbjerg and Thygesen}(2010)}]{Kaasbjerg10085102}
\bibinfo{author}{\bibnamefont{Kaasbjerg}} \bibnamefont{and}
  \bibinfo{author}{\bibfnamefont{K.~S.} \bibnamefont{Thygesen}},
  \bibinfo{journal}{Phys. Rev. B} \textbf{\bibinfo{volume}{81}},
  \bibinfo{pages}{085102} (\bibinfo{year}{2010}).

\bibitem[{\citenamefont{Tiago et~al.}(2008)\citenamefont{Tiago, Kent, Hood, and
  Reboredo}}]{Tiago08084311}
\bibinfo{author}{\bibfnamefont{M.~L.} \bibnamefont{Tiago}},
  \bibinfo{author}{\bibfnamefont{P.~R.~C.} \bibnamefont{Kent}},
  \bibinfo{author}{\bibfnamefont{R.~Q.} \bibnamefont{Hood}}, \bibnamefont{and}
  \bibinfo{author}{\bibfnamefont{F.~A.} \bibnamefont{Reboredo}},
  \bibinfo{journal}{J. Chem. Phys.} \textbf{\bibinfo{volume}{129}},
  \bibinfo{pages}{084311} (\bibinfo{year}{2008}).

\bibitem[{\citenamefont{Palummo et~al.}(2009)\citenamefont{Palummo, Hogan,
  Sottile, Bagala, and Rubio}}]{Palummo09084102}
\bibinfo{author}{\bibfnamefont{M.}~\bibnamefont{Palummo}},
  \bibinfo{author}{\bibfnamefont{C.}~\bibnamefont{Hogan}},
  \bibinfo{author}{\bibfnamefont{F.}~\bibnamefont{Sottile}},
  \bibinfo{author}{\bibfnamefont{P.}~\bibnamefont{Bagala}}, \bibnamefont{and}
  \bibinfo{author}{\bibfnamefont{A.}~\bibnamefont{Rubio}}, \bibinfo{journal}{J.
  Chem. Phys.} \textbf{\bibinfo{volume}{131}}, \bibinfo{pages}{084102}
  (\bibinfo{year}{2009}).

\bibitem[{\citenamefont{Rojas et~al.}(1995)\citenamefont{Rojas, Godby, and
  Needs}}]{Rojas951827}
\bibinfo{author}{\bibfnamefont{H.~N.} \bibnamefont{Rojas}},
  \bibinfo{author}{\bibfnamefont{R.~W.} \bibnamefont{Godby}}, \bibnamefont{and}
  \bibinfo{author}{\bibfnamefont{R.~J.} \bibnamefont{Needs}},
  \bibinfo{journal}{Phys. Rev. Lett.} \textbf{\bibinfo{volume}{74}},
  \bibinfo{pages}{1827} (\bibinfo{year}{1995}).

\bibitem[{\citenamefont{Rieger et~al.}(1999)\citenamefont{Rieger, Steinbeck,
  White, Rojas, and Godby}}]{Rieger99211}
\bibinfo{author}{\bibfnamefont{M.~M.} \bibnamefont{Rieger}},
  \bibinfo{author}{\bibfnamefont{L.}~\bibnamefont{Steinbeck}},
  \bibinfo{author}{\bibfnamefont{I.~D.} \bibnamefont{White}},
  \bibinfo{author}{\bibfnamefont{H.~N.} \bibnamefont{Rojas}}, \bibnamefont{and}
  \bibinfo{author}{\bibfnamefont{R.~W.} \bibnamefont{Godby}},
  \bibinfo{journal}{Comput. Phys. Commun.} \textbf{\bibinfo{volume}{117}},
  \bibinfo{pages}{211 } (\bibinfo{year}{1999}).

\bibitem[{\citenamefont{Faleev et~al.}(2004)\citenamefont{Faleev, van
  Schilfgaarde, and Kotani}}]{Faleev04126406}
\bibinfo{author}{\bibfnamefont{S.~V.} \bibnamefont{Faleev}},
  \bibinfo{author}{\bibfnamefont{M.}~\bibnamefont{van Schilfgaarde}},
  \bibnamefont{and} \bibinfo{author}{\bibfnamefont{T.}~\bibnamefont{Kotani}},
  \bibinfo{journal}{Phys. Rev. Lett.} \textbf{\bibinfo{volume}{93}},
  \bibinfo{pages}{126406} (\bibinfo{year}{2004}).

\bibitem[{NIS()}]{NIST}
\bibinfo{note}{NIST Computational Chemistry Comparison and Benchmark Database,
  http://cccbdb.nist.gov}.

\bibitem[{NWC()}]{NWChem}
\bibinfo{note}{NWChem, A Computational Chemistry Package for Parallel
  Computers, Pacific Northwest National Lab, Richland, Washington, USA (2003).}

\end{thebibliography}

\end{document}